\documentclass[]{spie}

\usepackage{amsmath,amsfonts,amssymb}
\usepackage{graphicx}
\usepackage[acronym]{glossaries}
\usepackage{aas_macros}

\makeglossaries
\newacronym{adi}{ADI}{angular differential imaging}
\newacronym{cnn}{CNN}{Convolutional Neural Network}

\usepackage[colorlinks=true, allcolors=blue]{hyperref}

\title{Machine learning revolution for exoplanet direct imaging detection: transformer architectures}

\author[a]{Yu-Chia Lin}
\affil[a]{University of Arizona, Tucson, AZ, USA}

\authorinfo{Further author information: (Send correspondence to Y.-C. L.)\\Y.-C. L.: E-mail: yuchialin@arizona.edu}

\begin{document} 
\maketitle

\begin{abstract}
Directly imaging exoplanets is a formidable challenge due to extreme contrast ratios and quasi-static speckle noise, motivating the exploration of advanced post-processing methods. While \glspl{cnn} have shown promise, their inherent limitations in capturing long-range dependencies in image sequences hinder their effectiveness. This study introduces a novel hybrid deep learning architecture that combines a \gls{cnn} feature extractor with a Transformer encoder to leverage temporal information, modeling the signature of a planet's coherent motion across an observation sequence. We first validated the model on a purely synthetic dataset, where it demonstrated excellent performance. While the final metrics varied slightly between training runs, our reported trial achieved 100.0\% accuracy, a 100.0\% F1-score, and a position accuracy of 0.72 pixels, showing strong results on this specific test case in comparison to traditional methods like median subtraction and PCA-KLIP. To assess its viability on realistic data, we retrained the model on a semi-synthetic dataset created by injecting planet signals into actual high-contrast imaging observations of the TW Hya protoplanetary disk from JWST. The model successfully identified the injected signals with high confidence, confirming its ability to function amidst complex, correlated noise and bright disk features. This work serves as a successful proof-of-concept, demonstrating that a \gls{cnn}-Transformer architecture holds significant promise as a fast, accurate, and automated method for exoplanet detection in the large datasets expected from current and future high-contrast imaging instruments.
\end{abstract}


\section{INTRODUCTION}
\label{sec:intro}

The direct imaging of exoplanets is one of the most challenging frontiers in astronomy. The primary difficulty lies in detecting the faint signal of a planet buried in the overwhelming glare of its host star, with contrast ratios often exceeding $10^6$ \cite{2023ASPC..534..799C}. Telescopes equipped with coronagraphs and adaptive optics systems are designed to suppress starlight, but residual light from quasi-static speckles remains a significant source of noise.

Conventional post-processing techniques have been developed to mitigate this noise. For example, PCA-KLIP is an algorithm based on Principal Component Analysis (PCA) that is widely used in high-contrast imaging \cite{2015ascl.soft06001W}. While these methods have been instrumental in many discoveries, the increasing volume and complexity of data from new instruments create an opportunity for complementary approaches. Techniques like machine learning can offer advantages in processing speed and the potential to learn complex patterns that are difficult to model analytically.

In recent years, \glspl{cnn} have been successfully applied to direct imaging of exoplanets, enabling planet detection with fast inference times and without prior background estimation. A notable example is \citenum{2023SPIE12680E..28A}, which utilized a \gls{cnn} to achieve high-precision detection on simulated data. However, standard \gls{cnn} architectures possess inherent limitations; their localized receptive fields make it difficult to capture global features, and their translation-invariant nature can discard important temporal or sequential information.

To address these shortcomings, this paper presents a preliminary investigation into a hybrid \gls{cnn}-Transformer model. Transformers, first introduced by \citenum{2017arXiv170603762V}, use a self-attention mechanism that has revolutionized natural language processing by effectively modeling long-range dependencies in sequential data. The self-attention mechanism is uniquely suited for this task, as it can weigh the importance of different frames in a sequence, allowing it to learn the signature of coherent motion while ignoring stochastic or static noise. It can thus differentiate the temporal signature of a faint, moving planet from that of a bright, quasi-static speckle. We adapt this capability to analyze sequences of high-contrast images, treating them as a time series to distinguish the faint, coherent motion of a planet from uncorrelated background noise. The sequence of images can be regarded not only as time-series images, but could also represent the sequence of images obtained by rotating the telescope. We detail the model's architecture, training, and performance on both simulated and real astronomical data.

\section{METHODOLOGY}
\label{sec:methodology}

Our approach involves a supervised deep learning model trained on simulated high-contrast imaging sequences. We compare its performance against several traditional algorithms on a standardized test set to demonstrate its potential. An example Jupyter notebook of this demo code is available online, \url{https://github.com/dreamjade/CNN_Transformer_ExoPlanet/blob/main/CNN_Transformer.ipynb}.

\subsection{Simulated Datasets}
To validate our model, we designed two distinct datasets. The first is a purely synthetic dataset to establish baseline performance and validate the model's core logic. The second is a semi-synthetic dataset, using real on-sky data, to assess the model's viability under more realistic noise conditions.

\subsubsection{Synthetic Dataset for Model Validation}
This initial dataset was designed to test the model's fundamental ability to detect a faint, moving signal within a noisy background. This simulation is simplified and does not fully represent the complexity of high-contrast imaging data. Each sample consists of a sequence of 10 frames, each 64x64 pixels in size. The dataset is balanced, with a 50\% probability for any given sequence to contain a moving signal.

\begin{itemize}
    \item \textbf{Static Noise Sequences:} The background image for these sequences is Gaussian random noise drawn from a distribution with a mean of 0 and a standard deviation equal to the `noise\_level` hyperparameter (set to 0.3). We then add some stationary bright spots with randomized brightness levels between 0.1 and 0.3 to act as distractors, analogous to static speckles.
    
    \item \textbf{Moving Dot Sequences:} For sequences containing a target, we add a 3-pixel dot to the static noise background. The dot moves along a circular orbit and has a fixed peak brightness of 0.5. The background noise has a standard deviation of `noise\_level` = 0.3, resulting in a signal-to-noise ratio (SNR) of approximately 1.67 ($0.5 / 0.3$) for a single image in the sequences.
\end{itemize}

\begin{figure}[ht]
\begin{center}
\includegraphics[width=0.9\textwidth]{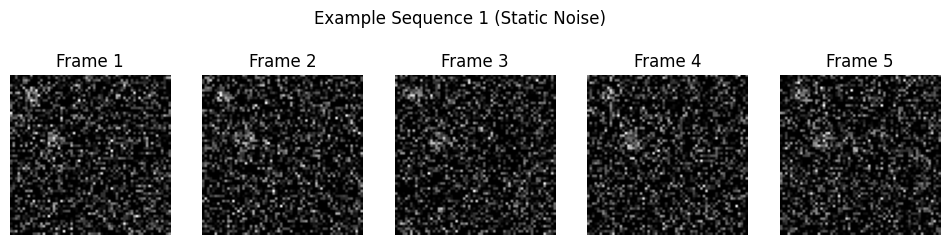}
\vspace{5mm}
\includegraphics[width=0.9\textwidth]{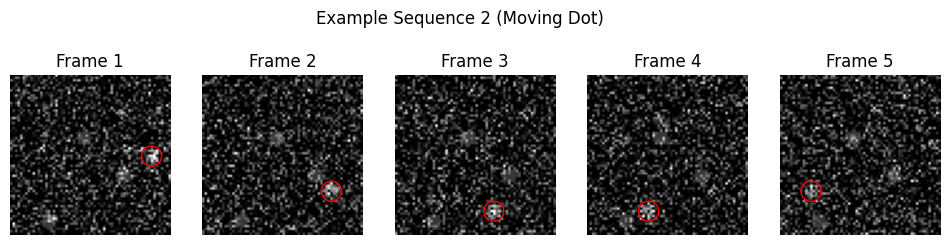}
\end{center}
\caption[]
{ \label{fig:synth_seq}
Example sequences from the synthetic dataset. The top panel shows a "Static Noise" sequence, which contains only a noisy background with bright, stationary distractors and no coherent moving signal. The bottom panel shows a "Moving Dot" sequence, where a simulated planet (indicated by the red circle) moves along its orbit over five frames. The model is trained to distinguish between these two classes and the planet location in a "Moving Dot" sequence.}
\end{figure}

The final dataset was split into 2,000 training and 500 test samples. Examples of both "Moving Dot" and "Static Noise" sequences are shown in Fig.~\ref{fig:synth_seq}.

\subsubsection{Semi-Synthetic On-Sky Dataset}
To evaluate the model's performance under more realistic conditions, we created a semi-synthetic dataset using actual high-contrast imaging observations of the TW Hya system from \citenum{lin_in_prep} via JWST. This approach preserves the complex, correlated noise structures and bright features of the protoplanetary disk.

\begin{itemize}
    \item \textbf{Base Data:} We used a real astronomical frame of TW Hya from JWST as the background.
    \item \textbf{Planet Injection:} A synthetic planet signal, modeled as a 3-pixel Gaussian dot, was injected into each frame of the sequence. The planet follows a circular orbit. The brightness of the injected signal was set to achieve an SNR of 5, calculated relative to the standard deviation of the flux within an annular region at the same orbital radius as the injected planet. Poisson noise was also applied to the frames to simulate photon counting statistics.
    \item \textbf{Dataset Creation:} As with the synthetic data, we generated sequences with and without injected planets to create a balanced dataset for training and testing.
\end{itemize}

\subsection{Model Architecture}
\begin{figure}[ht]
   \begin{center}
   \includegraphics[width=1.0\textwidth]{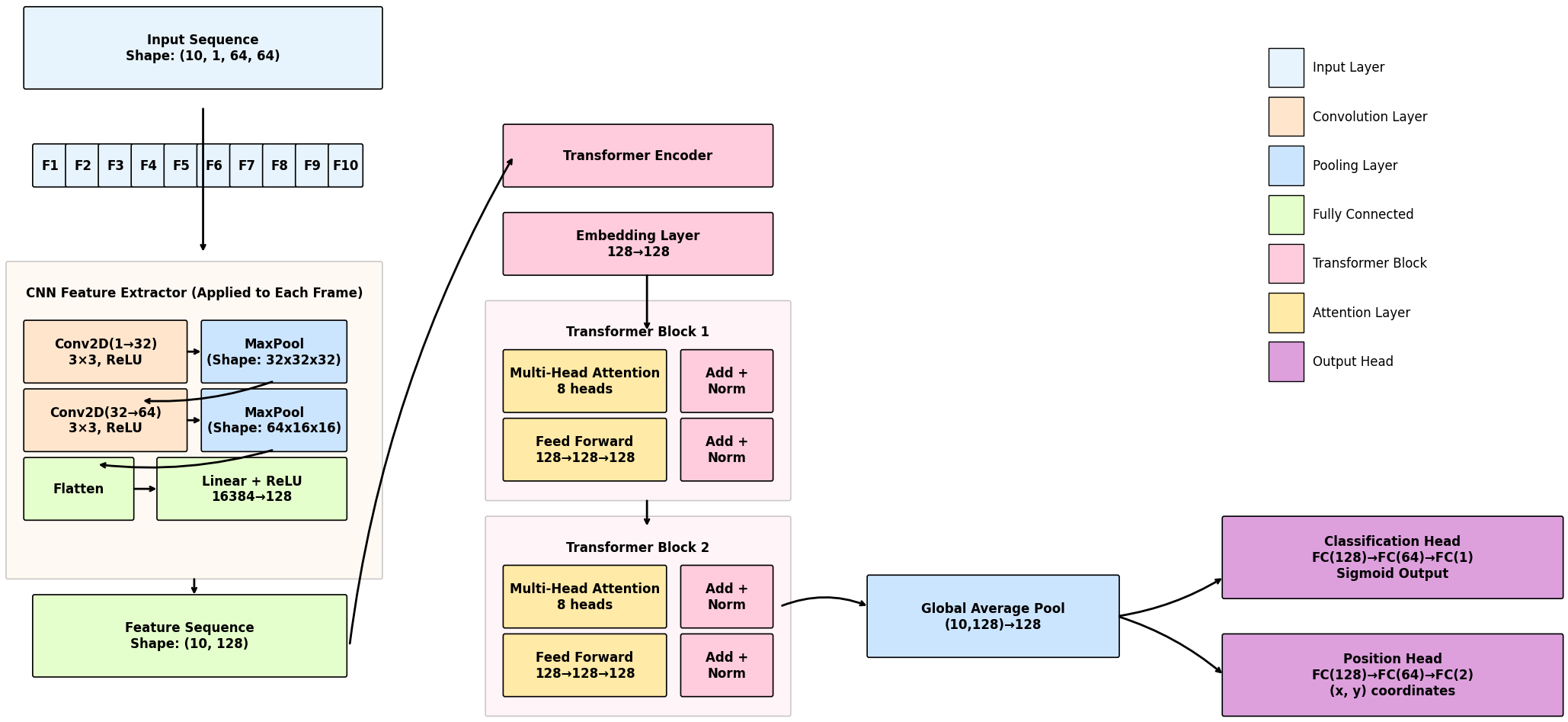}
   \end{center}
   \caption[architecture] 
   { \label{fig:arch} 
   Diagram of the hybrid \gls{cnn}-Transformer architecture. An input sequence of 10 frames is processed by a \gls{cnn} feature extractor. The resulting sequence of feature vectors is then passed to a two-block Transformer Encoder, which uses self-attention to model temporal relationships. The final output is split into a classification head, which predicts the presence of a planet, and a position head, which predicts its coordinates.}
   \end{figure}

Our proposed model is a hybrid architecture combining a \gls{cnn} feature extractor with a Transformer Encoder, as illustrated in Fig.~\ref{fig:arch}.

\begin{itemize}
    \item \textbf{\gls{cnn} Feature Extractor:} Each frame in the input sequence is processed independently by a \gls{cnn}. The \gls{cnn} consists of two convolutional layers with 3x3 kernels and ReLU activations, each followed by a max-pooling layer. This network reduces the dimensionality of each image and extracts salient features, producing a 128-element feature vector for each frame.
    \item \textbf{Transformer Encoder:} The sequence of feature vectors from the \gls{cnn} is passed to a Transformer Encoder. The encoder consists of two Transformer blocks, each containing a multi-head self-attention layer (with 8 heads) and a feed-forward network. This architecture allows the model to weigh the importance of different frames in the sequence and identify temporal patterns characteristic of a moving planet.
    \item \textbf{Output Heads:} The output from the Transformer is fed into two separate fully-connected heads:
    \begin{enumerate}
        \item A \textbf{Classification Head} with a sigmoid activation function that outputs a probability (0 to 1) of a moving object being present.
        \item A \textbf{Position Head} that performs regression to predict the (x, y) coordinates of the detected object in the first frame.
    \end{enumerate}
\end{itemize}

\subsection{Training}
The model was trained for 50 or 100 epochs using the Adam optimizer with a learning rate of $10^{-4}$ and a batch size of 32 for the Simulated Datasets and the Semi-Synthetic On-Sky Dataset, respectively. A combined loss function was employed, using a 10:1 weighting for the Binary Cross-Entropy (classification) and Position Error (Euclidean distance) components.

\subsection{Comparison Methods}
To benchmark our model's performance in this demonstration, we implemented several standard algorithms used in high-contrast imaging:
\begin{itemize}
    \item \textbf{Median Subtraction:} The median of all frames in the sequence is subtracted from each frame to remove static background features.
    \item \textbf{Optimized Median Subtraction:} An enhanced version of median subtraction that incorporates a confidence-based threshold to optimize detection accuracy.
    \item \textbf{PCA-KLIP:} A method that uses Principal Component Analysis to model and subtract the stellar PSF and associated speckle noise \cite{2015ascl.soft06001W}.
\end{itemize}

\section{RESULTS AND ANALYSIS}
\label{sec:results}

\subsection{Synthetic Data Test}
\begin{figure}[ht]
   \begin{center}
   \includegraphics[width=1.0\textwidth]{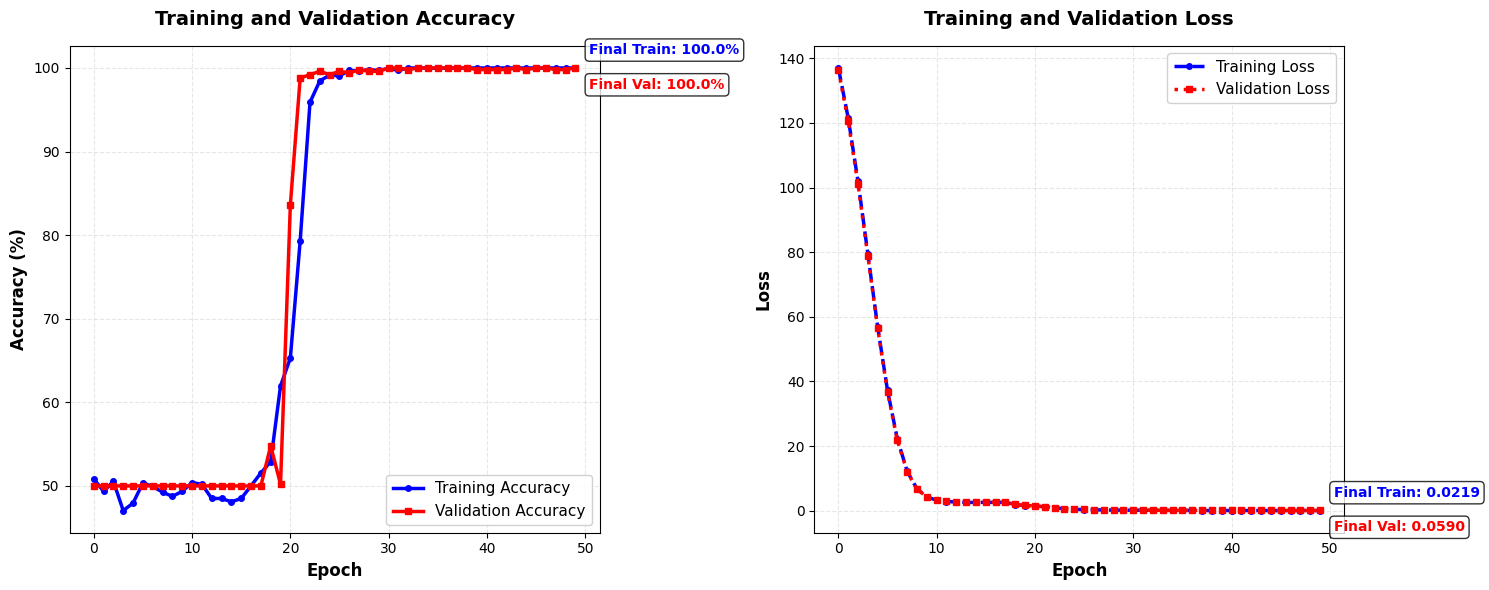}
   \end{center}
   \caption[training-curves] 
   { \label{fig:training_curves} 
   Training and validation curves for the model trained on the synthetic dataset. The left panel shows accuracy, and the right panel shows loss, both as a function of training epoch. The model from this training run achieves 100.0\% final accuracy on both the training and validation sets, with the loss converging to a low value, indicating effective learning.}
   \end{figure}

The model was first trained and evaluated on the purely synthetic dataset. The training and validation curves (Fig.~\ref{fig:training_curves}) show that the model rapidly converges.

Due to the stochastic nature of the training process, final accuracies typically reached near-perfect levels ($\geq99\%$), with the representative trial presented here achieving 100.0\% accuracy and low loss on both the training and validation sets. This indicates robust learning for this simplified test case without significant overfitting.

\begin{figure}[ht]
   \begin{center}
   \includegraphics[width=1.0\textwidth]{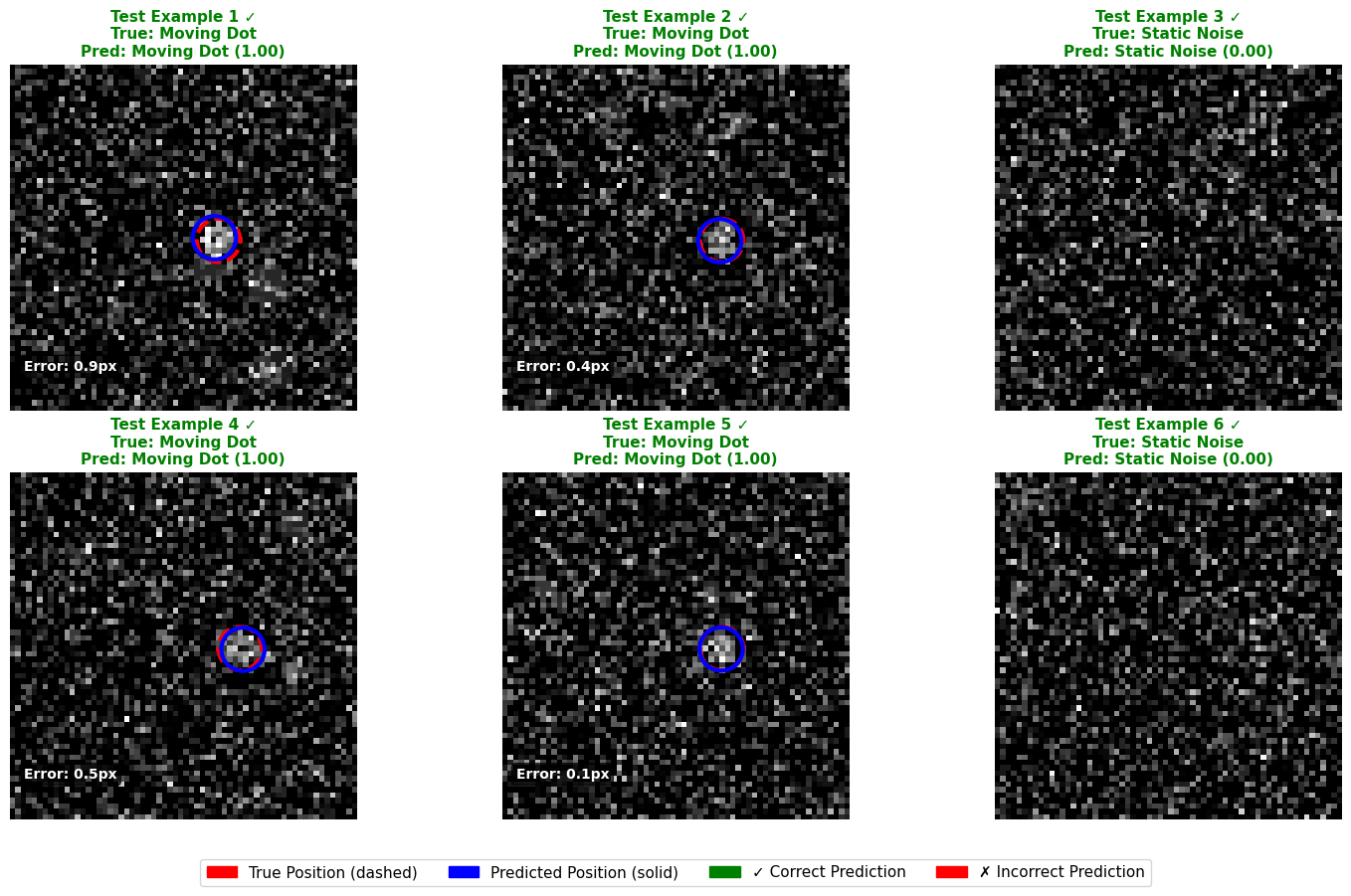}
   \end{center}
   \caption[] 
   { \label{fig:synth_preds} 
   Model predictions on six test examples from the synthetic dataset. The model correctly classifies all four “Moving Dot” sequences and both “Static Noise” sequences with high confidence. For the detected signals, the predicted position (solid blue circle) is shown alongside the true position (dashed red circle), with sub-pixel position errors noted in each case. The results demonstrate the model's excellent accuracy in both classification and position estimation.}
   \end{figure}

   Figure~\ref{fig:synth_preds} demonstrates the model's performance on six examples from the test set. It correctly classifies all sequences containing moving dots and those with only static noise, achieving perfect confidence scores in each case. For the detected signals, the position error—the Euclidean distance between the true and predicted coordinates—is consistently sub-pixel, highlighting the model's high precision.

   To quantitatively evaluate the classification performance, we use standard statistical metrics. A True Positive (TP) is a correct identification of a planet, a True Negative (TN) is a correct identification of a sequence with no planet, a False Positive (FP) is an incorrect identification of a planet where none exists, and a False Negative (FN) is a failure to detect a planet that is present. From these, we define the following metrics:
    \begin{equation}
    \text{Accuracy} = \frac{\text{TP} + \text{TN}}{\text{TP} + \text{TN} + \text{FP} + \text{FN}}
    \end{equation}
    \begin{equation}
    \text{Precision} = \frac{\text{TP}}{\text{TP} + \text{FP}}
    \end{equation}
    \begin{equation}
    \text{Recall} = \frac{\text{TP}}{\text{TP} + \text{FN}}
    \end{equation}
    \begin{equation}
    \text{F1-Score} = 2 \times \frac{\text{Precision} \times \text{Recall}}{\text{Precision} + \text{Recall}}
    \end{equation}
   The F1-score provides a balanced measure of a model's performance by taking the harmonic mean of precision and recall.

   Table~\ref{tab:comparison} compares the results from all tested methods on this controlled dataset. The \gls{cnn}-Transformer model achieved high performance across all metrics. For instance, while PCA-KLIP demonstrated excellent position accuracy for its detections, its performance on this specific test set, which has a limited number of frames, is reflected in a lower recall rate and F1-score. The comparatively lower recall of PCA-KLIP can be attributed to the limited number of frames, which provides a smaller basis for constructing the principal components.

\begin{table}[ht]
\caption{Comparison of detection methods on the synthetic dataset.} 
\label{tab:comparison}
\begin{center}       
\begin{tabular}{|l|c|c|c|c|c|}
\hline
\rule[-1ex]{0pt}{3.5ex} \textbf{Method} & \textbf{F1-Score} & \textbf{Accuracy} & \textbf{Precision} & \textbf{Recall} & \textbf{Pos Error (pixels)} \\
\hline
\rule[-1ex]{0pt}{3.5ex} \textbf{\gls{cnn}-Transformer} & \textbf{100.00\%} & \textbf{100.00\%} & \textbf{100.00\%} & \textbf{100.00\%} & \textbf{0.72} \\
\hline
\rule[-1ex]{0pt}{3.5ex} Optimized Median Sub & 89.39\% & 89.60\% & 91.25\% & 87.60\% & 1.75 \\
\hline
\rule[-1ex]{0pt}{3.5ex} PCA-KLIP & 52.17\% & 70.67\% & 100.00\% & 35.29\% & 1.45 \\
\hline
\end{tabular}
\end{center}
\end{table} 

\subsection{On-Sky Data Application: TW Hya}
To assess the model's viability for real-world applications, we applied it to the semi-synthetic on-sky dataset of the protoplanetary disk around the young T Tauri star, TW Hya. We used the data from JWST \cite{lin_in_prep}. This star, located at a distance of 60.14 $\ pm$0.05 pc \cite{2023A&A...674A...1G}, is approximately $8 \pm 3$ Myr old \cite{2018ApJ...853..120S} and is chosen due to its nearly face-on orientation.

\begin{figure}[ht]
   \begin{center}
   \includegraphics[width=1.0\textwidth]{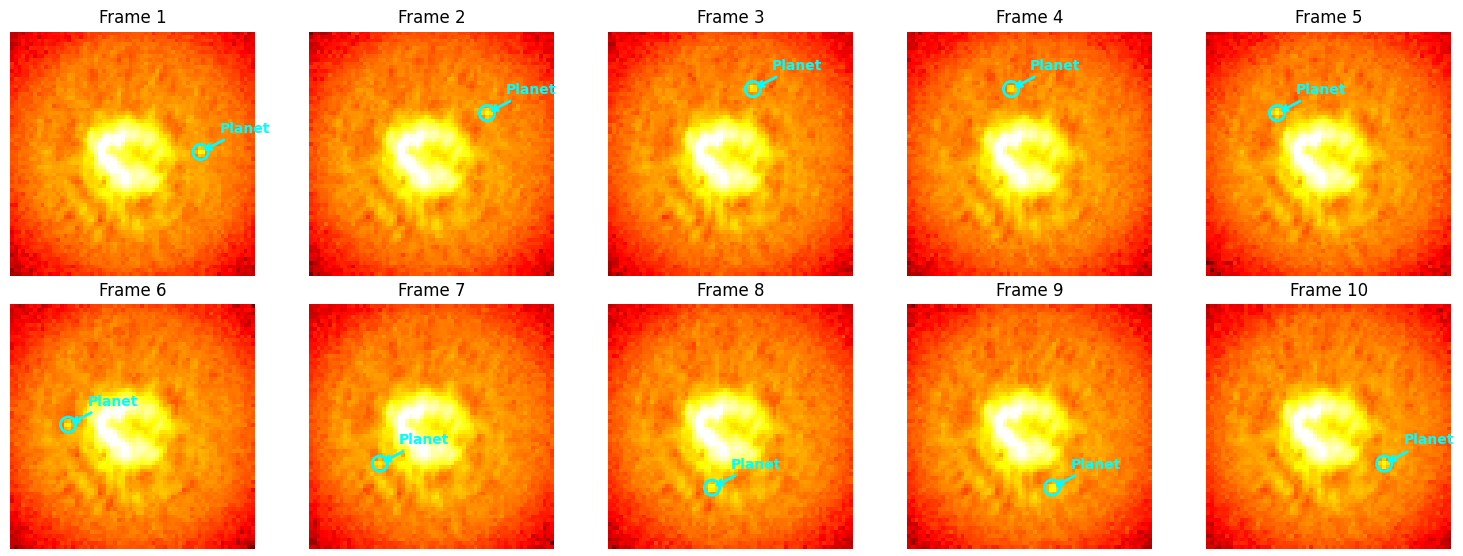}
   \end{center}
   \caption[tw-hya-seq] 
   { \label{fig:tw_hya_seq} 
   An example sequence from the semi-synthetic on-sky dataset. A synthetic planet signal, indicated by the cyan markers, is injected into a sequence of real JWST observations of TW Hya. This method preserves the complex noise and background structures from the protoplanetary disk.}
   \end{figure}

\begin{figure}[ht]
   \begin{center}
   \includegraphics[width=1.0\textwidth]{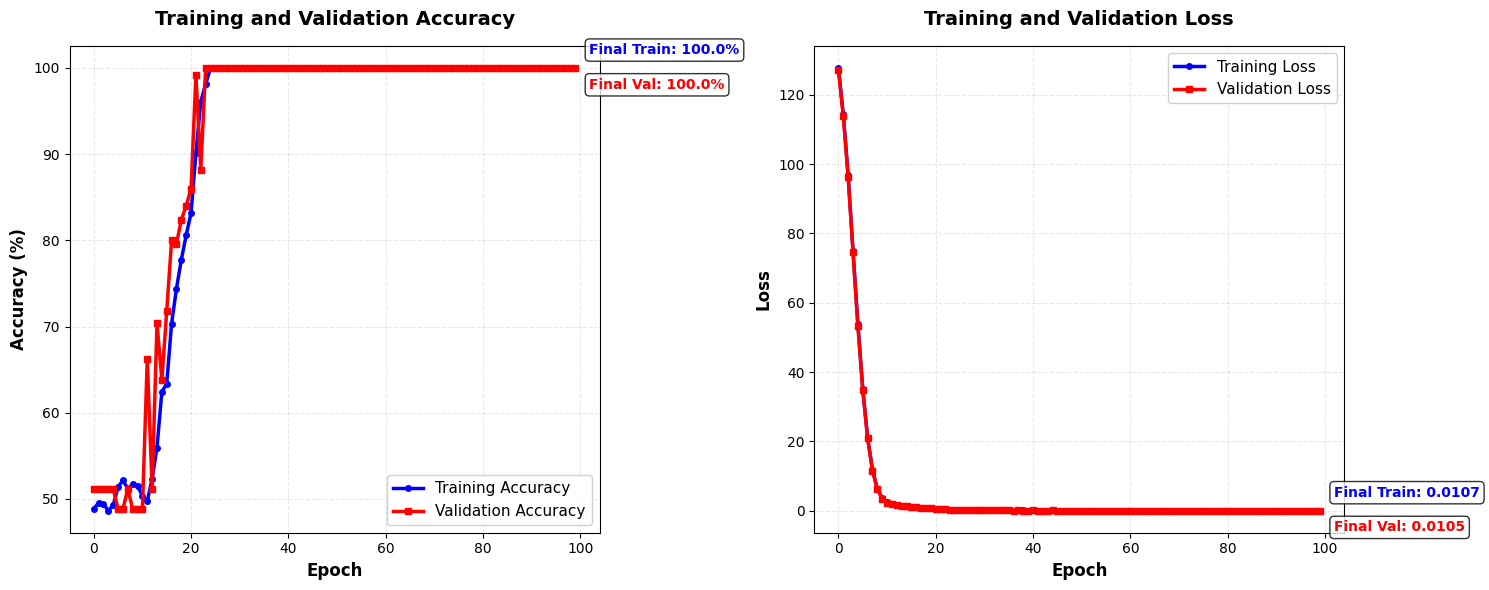}
   \end{center}
   \caption[tw-hya-curves] 
   { \label{fig:tw_hya_curves} 
   Training and validation curves for the model trained on the semi-synthetic TW Hya dataset. The model again reaches 100.0\% accuracy and low loss on both training and validation sets, demonstrating its ability to learn the planet's signature amidst the complex structure of a real protoplanetary disk.}
   \end{figure}

We used a sequence of observations of TW Hya and injected a synthetic planet signal with a known trajectory, as shown in the example in Fig.~\ref{fig:tw_hya_seq}. The model was then trained on this semi-synthetic data. As shown in Fig.~\ref{fig:tw_hya_curves}, the model again achieved 100\% training and validation accuracy, demonstrating its ability to learn the planet's signature amidst the complex, correlated noise structures and bright features of a real protoplanetary disk.

\begin{figure}[ht]
   \begin{center}
    \includegraphics[width=1.0\textwidth]{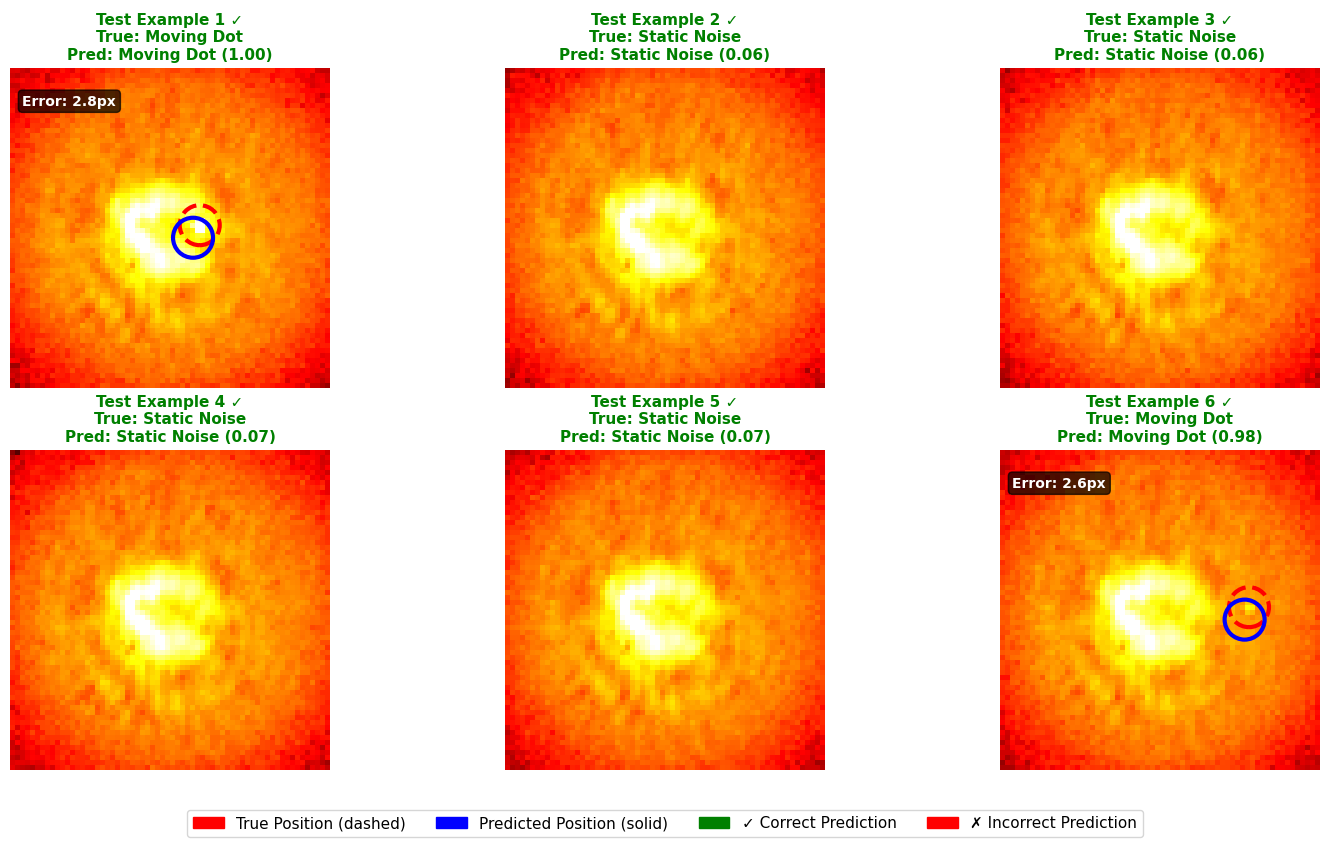}
   \end{center}
   \caption[tw-hya-preds] 
   { \label{fig:tw_hya_preds} 
   Model prediction on a test sequence from the semi-synthetic TW Hya dataset. The red circle indicates the true position of the injected planet in each frame. In the first frame, the model's successful prediction is shown with a blue circle and yellow text, demonstrating a position accuracy of a few pixels in a realistic data environment.}
   \end{figure}

   The model's predictions on the test set of injected TW Hya data are shown in Fig.~\ref{fig:tw_hya_preds}. The model successfully identifies the injected planet in all "Moving Dot" cases with high confidence and achieves a position accuracy of a few pixels, confirming its effectiveness on realistic astronomical data.

   \subsection{Limitations of Preliminary Results}
   While the model achieved excellent performance on our test cases, these results should be interpreted as a successful proof-of-concept that validates the architecture's potential. The datasets used in this preliminary study represent simplified scenarios with circular orbits and relatively high-SNR planets. Furthermore, the model was retrained for the semi-synthetic on-sky dataset, indicating that the current approach requires specific training for different noise characteristics and observational conditions. Developing a single, generalized model applicable across various instruments and targets without retraining will require a much larger and more diverse training dataset. This study serves as a foundational step, demonstrating the architecture's core capability before its application to more complex, archival datasets.

\section{FUTURE WORK}
\label{sec:future}
This work demonstrates the powerful potential of combining \glspl{cnn} and Transformers for exoplanet detection. As this is a preliminary study, our future research will focus on several key areas to enhance the model's capabilities and move from a demonstration to a fully validated tool:
\begin{itemize}
    \item \textbf{Robustness and Generalization:} Our immediate next step is to test the model against more challenging and realistic datasets. This includes injecting planets with a wider range of signal-to-noise ratios, eccentric orbits, and testing on data with more complex, correlated speckle noise patterns.
    \item \textbf{Meta-Learning for Instrumental Adaptation:} To create a truly generalizable tool, we plan to develop a meta-learning framework that can adapt the model to different instruments by providing it with instrumental Point Spread Functions (PSFs) or other observational parameters as input.
    \item \textbf{Incorporating Prior Information:} The model's performance could be further improved by incorporating prior information from other detection methods, such as potential orbits and planet sizes derived from transit surveys.
    \item \textbf{Multi-Frequency Data:} We will explore architectures capable of fusing data from different frequencies, which could help in distinguishing planetary signals from stellar artifacts based on their spectral characteristics.
    \item \textbf{Order-Free Models:} We will investigate order-free Transformer architectures, like \citenum{2017arXiv170705495C} and \citenum{2019arXiv190903434T}, which have shown success in multi-label image classification, to potentially detect multiple planets within the same image sequence.
\end{itemize}

\section{CONCLUSION}
\label{sec:conclusion}
The key advantage of this architecture lies in the Transformer's self-attention mechanism, which enables the model to learn the temporal signatures of orbiting planets directly from pixel data. This work represents a methodological shift from static image analysis or engineered feature extraction toward a sequence-based, deep learning approach. As the volume of data from instruments like JWST and future observatories grows, such fast and automated techniques will be crucial for efficiently vetting candidates and accelerating discovery.

\acknowledgments       
This work is based in part on observations made with the NASA/ESA/CSA James Webb Space Telescope.
The data were obtained from the Mikulski Archive for Space Telescopes at the Space Telescope Science Institute, which is operated by the Association of Universities for Research in Astronomy, Inc., under NASA contract NAS 5-03127 for JWST. These observations are associated with program 1179. Portions of this research were supported by funding from the Technology Research Initiative Fund (TRIF) of the Arizona Board of Regents and by generous anonymous philanthropic donations to the Steward Observatory of the College of Science at the University of Arizona. The authors thank the SPIE organizing committee for the opportunity to present this work. I also acknowledge the use of AI tools, GitHub Copilot for code documentation, and Google AI Studio for proceeding readability improvements.

\bibliography{report}

\begin{thebibliography}{1}

\bibitem{2023ASPC..534..799C}
{Currie}, T., {Biller}, B., {Lagrange}, A., {Marois}, C., {Guyon}, O., {Nielsen}, E.~L., {Bonnefoy}, M., and {De Rosa}, R.~J., ``{Direct Imaging and Spectroscopy of Extrasolar Planets},'' in [{\em Protostars and Planets VII}{\nolinebreak\hspace{0.1em}]},  {Inutsuka}, S., {Aikawa}, Y., {Muto}, T., {Tomida}, K., and {Tamura}, M., eds., {\em Astronomical Society of the Pacific Conference Series} {\bf 534},  799 (July 2023).

\bibitem{2015ascl.soft06001W}
{Wang}, J.~J., {Ruffio}, J.-B., {De Rosa}, R.~J., {Aguilar}, J., {Wolff}, S.~G., and {Pueyo}, L., ``{pyKLIP: PSF Subtraction for Exoplanets and Disks}.'' Astrophysics Source Code Library, record ascl:1506.001 (June 2015).

\bibitem{2023SPIE12680E..28A}
{Ahmed}, Z., {D'Amico}, S., {Hu}, R., and {Damiano}, M., ``{Exoplanet detection from starshade images using convolutional neural networks},'' in [{\em Society of Photo-Optical Instrumentation Engineers (SPIE) Conference Series}{\nolinebreak\hspace{0.1em}]},  {\em Society of Photo-Optical Instrumentation Engineers (SPIE) Conference Series} {\bf 12680},  1268028 (Oct. 2023).

\bibitem{2017arXiv170603762V}
{Vaswani}, A., {Shazeer}, N., {Parmar}, N., {Uszkoreit}, J., {Jones}, L., {Gomez}, A.~N., {Kaiser}, L., and {Polosukhin}, I., ``{Attention Is All You Need},'' {\em arXiv e-prints} ,  arXiv:1706.03762 (June 2017).

\bibitem{lin_in_prep}
Lin, Y.-C., Leisenring, J., Wolff, S.~G., Hom, J., and Douglas, E.~S., ``Jwst/nircam imaging of young stellar objects. iv. detailed imaging of the protoplanetary disk around tw hya,'' (2025, in preparation).

\bibitem{2023A&A...674A...1G}
{Gaia Collaboration}, {Vallenari}, A., {Brown}, A.~G.~A., {Prusti}, T., {de Bruijne}, J.~H.~J., {Arenou}, F., {Babusiaux}, C., {Biermann}, M., {Creevey}, O.~L., {Ducourant}, C., {Evans}, D.~W., {Eyer}, L., {Guerra}, R., {Hutton}, A., {Jordi}, C., {Klioner}, S.~A., {Lammers}, U.~L., {Lindegren}, L., {Luri}, X., {Mignard}, F., {Panem}, C., {Pourbaix}, D., {Randich}, S., {Sartoretti}, P., {Soubiran}, C., {Tanga}, P., {Walton}, N.~A., {Bailer-Jones}, C.~A.~L., {Bastian}, U., {Drimmel}, R., {Jansen}, F., {Katz}, D., {Lattanzi}, M.~G., {van Leeuwen}, F., {Bakker}, J., {Cacciari}, C., {Casta{\~n}eda}, J., {De Angeli}, F., {Fabricius}, C., {Fouesneau}, M., {Fr{\'e}mat}, Y., {Galluccio}, L., {Guerrier}, A., {Heiter}, U., {Masana}, E., {Messineo}, R., {Mowlavi}, N., {Nicolas}, C., {Nienartowicz}, K., {Pailler}, F., {Panuzzo}, P., {Riclet}, F., {Roux}, W., {Seabroke}, G.~M., {Sordo}, R., {Th{\'e}venin}, F., {Gracia-Abril}, G., {Portell}, J., {Teyssier}, D., {Altmann}, M., {Andrae}, R., {Audard}, M., {Bellas-Velidis},
  I., {Benson}, K., {Berthier}, J., {Blomme}, R., {Burgess}, P.~W., {Busonero}, D., {Busso}, G., {C{\'a}novas}, H., {Carry}, B., {Cellino}, A., {Cheek}, N., {Clementini}, G., {Damerdji}, Y., {Davidson}, M., {de Teodoro}, P., {Nu{\~n}ez Campos}, M., {Delchambre}, L., {Dell'Oro}, A., {Esquej}, P., {Fern{\'a}ndez-Hern{\'a}ndez}, J., {Fraile}, E., {Garabato}, D., {Garc{\'\i}a-Lario}, P., {Gosset}, E., {Haigron}, R., {Halbwachs}, J.~L., {Hambly}, N.~C., {Harrison}, D.~L., {Hern{\'a}ndez}, J., {Hestroffer}, D., {Hodgkin}, S.~T., {Holl}, B., {Jan{\ss}en}, K., {Jevardat de Fombelle}, G., {Jordan}, S., {Krone-Martins}, A., {Lanzafame}, A.~C., {L{\"o}ffler}, W., {Marchal}, O., {Marrese}, P.~M., {Moitinho}, A., {Muinonen}, K., {Osborne}, P., {Pancino}, E., {Pauwels}, T., {Recio-Blanco}, A., {Reyl{\'e}}, C., {Riello}, M., {Rimoldini}, L., {Roegiers}, T., {Rybizki}, J., {Sarro}, L.~M., {Siopis}, C., {Smith}, M., {Sozzetti}, A., {Utrilla}, E., {van Leeuwen}, M., {Abbas}, U., {{\'A}brah{\'a}m}, P., {Abreu Aramburu}, A.,
  {Aerts}, C., {Aguado}, J.~J., {Ajaj}, M., {Aldea-Montero}, F., {Altavilla}, G., {{\'A}lvarez}, M.~A., {Alves}, J., {Anders}, F., {Anderson}, R.~I., {Anglada Varela}, E., {Antoja}, T., {Baines}, D., {Baker}, S.~G., {Balaguer-N{\'u}{\~n}ez}, L., {Balbinot}, E., {Balog}, Z., {Barache}, C., {Barbato}, D., {Barros}, M., {Barstow}, M.~A., {Bartolom{\'e}}, S., {Bassilana}, J.~L., {Bauchet}, N., {Becciani}, U., {Bellazzini}, M., {Berihuete}, A., {Bernet}, M., {Bertone}, S., {Bianchi}, L., {Binnenfeld}, A., {Blanco-Cuaresma}, S., {Blazere}, A., {Boch}, T., {Bombrun}, A., {Bossini}, D., {Bouquillon}, S., {Bragaglia}, A., {Bramante}, L., {Breedt}, E., {Bressan}, A., {Brouillet}, N., {Brugaletta}, E., {Bucciarelli}, B., {Burlacu}, A., {Butkevich}, A.~G., {Buzzi}, R., {Caffau}, E., {Cancelliere}, R., {Cantat-Gaudin}, T., {Carballo}, R., {Carlucci}, T., {Carnerero}, M.~I., {Carrasco}, J.~M., {Casamiquela}, L., {Castellani}, M., {Castro-Ginard}, A., {Chaoul}, L., {Charlot}, P., {Chemin}, L., {Chiaramida}, V., {Chiavassa},
  A., {Chornay}, N., {Comoretto}, G., {Contursi}, G., {Cooper}, W.~J., {Cornez}, T., {Cowell}, S., {Crifo}, F., {Cropper}, M., {Crosta}, M., {Crowley}, C., {Dafonte}, C., {Dapergolas}, A., {David}, M., {David}, P., {de Laverny}, P., {De Luise}, F., and {De March}, R., ``{Gaia Data Release 3. Summary of the content and survey properties},'' {\em \aap}~{\bf 674},  A1 (June 2023).

\bibitem{2018ApJ...853..120S}
{Sokal}, K.~R., {Deen}, C.~P., {Mace}, G.~N., {Lee}, J.-J., {Oh}, H., {Kim}, H., {Kidder}, B.~T., and {Jaffe}, D.~T., ``{Characterizing TW Hydra},'' {\em \apj}~{\bf 853},  120 (Feb. 2018).

\bibitem{2017arXiv170705495C}
{Chen}, S.-F., {Chen}, Y.-C., {Yeh}, C.-K., and {Wang}, Y.-C.~F., ``{Order-Free RNN with Visual Attention for Multi-Label Classification},'' {\em arXiv e-prints} ,  arXiv:1707.05495 (July 2017).

\bibitem{2019arXiv190903434T}
{Tsai}, C.-P. and {Lee}, H.-Y., ``{Order-free Learning Alleviating Exposure Bias in Multi-label Classification},'' {\em arXiv e-prints} ,  arXiv:1909.03434 (Sept. 2019).

\end{thebibliography}
\bibliographystyle{spiebib}

\end{document}